\documentclass[journal=jacsat,manuscript=article]{achemso}

\usepackage[version=3]{mhchem} 
\usepackage{xcolor}

\usepackage{graphicx}
\usepackage{color}
\usepackage{dcolumn} 
\usepackage{bm}      
\usepackage{amssymb}
\usepackage{amsfonts}
\usepackage{svg}
\usepackage{comment}
\usepackage{xr-hyper}
\usepackage{float}
\usepackage[bookmarks=false,colorlinks,citecolor=red]{hyperref}
\usepackage{caption}
\usepackage{amsmath}
\usepackage{mhchem}
\usepackage{subcaption}
\usepackage[colorlinks,citecolor=red]{hyperref}
\usepackage{multicol}
\usepackage{multirow}
\usepackage{algorithm}
\usepackage{algpseudocode}
\usepackage{rotating}
\SectionNumbersOn

\newcommand{\COMMENTED}[1]{}

\makeatletter
\newcommand*{\addFileDependency}[1]{
  \typeout{(#1)}
  \@addtofilelist{#1}
  \IfFileExists{#1}{}{\typeout{No file #1.}}
}
\makeatother

\newcommand*{\myexternaldocument}[1]{%
    \externaldocument{#1}%
    \addFileDependency{#1.tex}%
    \addFileDependency{#1.aux}%
}

\myexternaldocument{supporting}

\author{Gopal R. Iyer}
\email{gopal_iyer@brown.edu}
\author{Brenda M. Rubenstein}
\affiliation
{Department of Chemistry, Brown University, Providence, Rhode Island 02912, USA}
\email{brenda_rubenstein@brown.edu}

\title[]
{Atomistic Descriptor Optimization Using Complementary Euclidean and Geodesic Distance Information}

\begin{document}


%





\begin{abstract}
Descriptors are physically-inspired, symmetry-preserving schemes for representing atomistic systems that play a central role in the construction of models of potential energy surfaces. Although physical intuition can be flexibly encoded into descriptor schemes, they are generally ultimately guided only by the spatial or topological arrangement of atoms in the system. However, since interatomic potential models aim to capture the variation of the potential energy with respect to atomic configurations, it is conceivable that they would benefit from descriptor schemes that implicitly encode both structural and energetic information rather than structural information alone.
Therefore, we propose a novel approach for the optimization of descriptors based on encoding information about geodesic distances along potential energy manifolds into the hyperparameters of commonly used descriptor schemes. To accomplish this, we combine two ideas: (1) a differential-geometric approach for the fast estimation of approximate geodesic distances [Zhu et al. \textit{J. Chem. Phys.} 150, 164103 (2019)]; and (2) an information-theoretic evaluation metric - \textit{information imbalance} - for measuring the shared information between two distance measures [Glielmo et al. \textit{PNAS Nexus}, 1, 1 (2022)]. Using three example molecules - ethanol, malonaldehyde, and aspirin - from the MD22 dataset, we first show that Euclidean (in Cartesian coordinates) and geodesic distances are
inequivalent distance measures, indicating the need for updated ground-truth distance measures that go beyond the Euclidean (or, more broadly, spatial) distance. We then utilize a Bayesian optimization framework to show that descriptors (in this case, atom-centered symmetry functions) can be optimized to maximally express a certain type of distance information, such as Euclidean or geodesic information. We also show that modifying the Bayesian optimization algorithm to minimize a combined objective function - the sum of the descriptor$\leftrightarrow$Euclidean and descriptor$\leftrightarrow$geodesic information imbalances - can yield descriptors that not only optimally express both Euclidean and geodesic distance information simultaneously, but in fact resolve substantial disagreements between descriptors optimized to encode only one type of distance measure. We discuss the relevance of our approach to the design of more physically rich and informative descriptors that can encode useful, alternative information about molecular systems. 
\end{abstract}

\section{Introduction}
\label{sec:introduction}
Machine learning-based interatomic potentials (MLPs) have emerged as computationally tractable solutions to the problem of modeling atomistic systems with high accuracy at larger length- and longer time-scales than are typically accessible to direct \textit{ab initio} calculations. By training flexible functional forms, such as neural networks on large pre-existing collections of high-accuracy \textit{ab initio} data, MLPs have successfully modeled a very large number of atomistic systems with high fidelity.\cite{mlp_1, mlp_2, ml_potentials_review, long_range_interactions, equivariant_potentials}

Many MLP models are structured like standard machine learning models, which may be summarized as: Model = Representation + Architecture, where the architecture refers to the functional form (e.g., a neural network) that processes the representation of the system to predict the output. In the context of MLPs, the representation is usually a physics-inspired featurization of the atomic positions referred to as a \textit{descriptor}. Descriptors are typically chosen to be functions of atomic positions that are invariant to a global rotation, translation, and permutation of identical atoms. Additionally, they may be chosen to encode some aspect of the physics of the specific atomistic system being studied, such as local structural arrangements. An important variation on this theme are \textit{kernel methods} that estimate the relative energies of atomic configurations using featurized kernels of the structural similarity between them.\cite{mlp_1, covariant_kernels, random_features, alchemical_similarity}

The field of descriptor design is vast and an exhaustive summary of the various strategies merits an independent review in its own right; see, for example, Ref. \citenum{descriptors}. In this work, we limit our discussion to atom-centered symmetry function (ACSF) descriptors, although, as we note later, our approaches readily generalize to other descriptor formulations. ACSF descriptors are among the first representation schemes to be have been demonstrated successfully for atomistic neural network potentials.\cite{behler, behler_parrinello} As the name suggests, ACSF descriptors involve constructing atom-wise functions that encode the local neighborhood of each atom (labeled $i$) in terms of different local structures, such as its pair-wise separation from its neighboring atoms ($r_{ij}$), the angles it forms with its neighboring triplets ($\theta_{ijk}$), etc.\cite{behler} Typical implementations of atomistic neural networks, termed Behler–Parrinello neural networks, then involve passing these representations to several individual neural networks, each corresponding to a single chemical element contained in the system, the outputs of which are then added to give a prediction for the energy of the atomic configuration. Force calculations are performed by taking the derivative of the energy with respect to the atom positions indirectly through derivatives with respect to the descriptors.\cite{behler_parrinello}

The construction of atomistic descriptors helps to determine whether or not two distinct atomic configurations encode similar physics, either because they are related by symmetry, or because the interactions between their atomic components lead to similar expressions for the energy - the most commonly modeled quantity - or other observables. In this work, we ask the following question: \textit{Is it possible to encode information beyond structural features - such as the approximate intrinsic geometry of the potential energy surface - within a given descriptor scheme?} If a descriptor scheme can be made to express the rich geometric structure of the potential energy manifold, the atomic fingerprints generated by such a descriptor will then, in principle, be distributed according to structural \textit{and energetic} similarity, making such representations more amenable to accurate modeling via an interatomic potential. Including energetic information content in descriptors is of particular relevance for the efficient sampling of structures for the training of data-driven models of force fields\cite{mlp_1,ml_potentials_review} and other quantities. When training such models with the aim of representing reactive dynamics, it is often crucial that higher energy structures surrounding transition states and intermediates along reaction pathways are represented in the training data, yet Euclidean distances alone often do not directly provide information about the energetics or spacing of configurations along reaction coordinates. Several schemes for structure sampling along reaction coordinates have been proposed, both in the context of materials modeling\cite{imbalzano_ceriotti_sampling_2018, shyue_ping_ong_sampling} and biomolecular simulation,\cite{tiwary_review} with the latter strongly focusing on the sampling of energetically favorable regions (configurations) separated by low-probability structural transformations, nonetheless, gaps remain. The need to better sample along reaction coordinates also raises the important question of which physical quantity is the most informative for describing geometric information about the potential energy manifold given a limited number of known atomic configurations of interest.

Given the fundamental importance of distance measures in describing geometries, we begin by noting that distance may serve as a meaningful, readily interpretable candidate for such a physical quantity. Several distance metrics have been proposed for atomistic systems, including direct Euclidean distance between Cartesian coordinate representations (assuming prior rotational alignment, e.g., via the Kabsch algorithm) and kernel-based similarity,\cite{mlp_1, covariant_kernels} including combined structural/alchemical dissimilarity kernels.\cite{alchemical_similarity} Despite significant developments in this regard, most popular similarity metrics are fundamentally grounded in the spatial structure of atomic arrangements, largely due to the apparent lack of any other type of information that can be gleaned from the structure \textit{a priori}. For instance, when measuring the expressivity of ACSF and SOAP (smooth overlap of atomic positions) descriptors, Glielmo \textit{et al.} (Ref. \citenum{distance_measures}) used Euclidean distance as the ground-truth distance measure to assess whether ACSF descriptors were able to encode distance information to a sufficient degree.

Here, we argue that the approximate \textit{geodesic distance} between atomic configurations along a potential energy manifold - which, as we will note, can be computed relatively easily - is an underutilized, yet physically informative, distance measure that contains information that may be fundamentally inaccessible to purely structure-based distance measures. Geodesic distances have been used in the context of dimensionality reduction for unsupervised atomistic machine learning, for example, in the ISOMAP algorithm.\cite{ceriotti_unsupervised} For the computation of approximate geodesic distances, we rely on the differential geometric approach developed by Zhu \textit{et al.} in Ref. \citenum{geodesic_interpolation} in the context of reaction pathway interpolation. We then show that it is possible to design and optimize descriptors to be simultaneously capable of encoding the information contained in both Euclidean and geodesic distances, thereby resolving the information imbalance\cite{distance_measures} between these contradictory metrics via a single optimal representation. For this purpose, we utilize a Bayesian optimization algorithm to query and optimize the hyperparameters of three different families of ACSF descriptors (chosen to have different levels of expressivity). We demonstrate our approach on three molecular sub-datasets of the MD22 dataset - ethanol, malonaldehyde, and aspirin.

The rest of this work is organized as follows. We describe the details of our approach in Section \ref{sec:theory}, including a description of the geodesic distance interpolation method (Section \ref{sec:fast_approximation_geodesic_distances}), the details of the Bayesian optimization algorithm (Section \ref{sec:theory_bayesian_optimization}), and the information-theoretic objective function for the Bayesian optimization originally proposed in Ref. \citenum{distance_measures} (Section \ref{sec:theory_information_imbalance}). Further computational details are provided in Section \ref{sec:computational_details}. In Section \ref{sec:results_and_discussion}, we first demonstrate the inequivalence of Euclidean and geodesic distances, and then discuss the application of the Bayesian optimization approach to encode Euclidean and geodesic distance information into the ACSF descriptors constructed in Section \ref{sec:acsf_descriptors}. We conclude and present an outlook for this approach in Section \ref{sec:conclusions}.

\section{Theory}
\label{sec:theory}
\subsection{The Need for Alternative Ground-Truth Distance Measures}
\label{sec:theory_motivation}
As discussed in Section \ref{sec:introduction}, Euclidean distance (measured in Cartesian coordinates)\footnote{Hereinafter, unless specified otherwise, the term Euclidean distance is assumed to mean that it is measured in Cartesian coordinates.} as well as other, more intricate similarity measures have been devised and utilized to interpret the relative arrangement of distinct atomistic configurations in an energy landscape in order to enable more accurate and generalizable predictive models. However, given that we are attempting to model the energy, which is a unique manifestation of the underlying physics of the atoms and electrons in the system, it seems limiting to rely only on spatially-informed similarity/distance measures without any concrete (even approximate) energetic information. This is especially significant when attempting to sample atomic configurations that feature sufficient structural and energetic diversity for potential models to be able to model both low- and high-energy configurations accurately. 

\begin{figure}[]
    \centering
    {\includegraphics[width=0.9\textwidth]{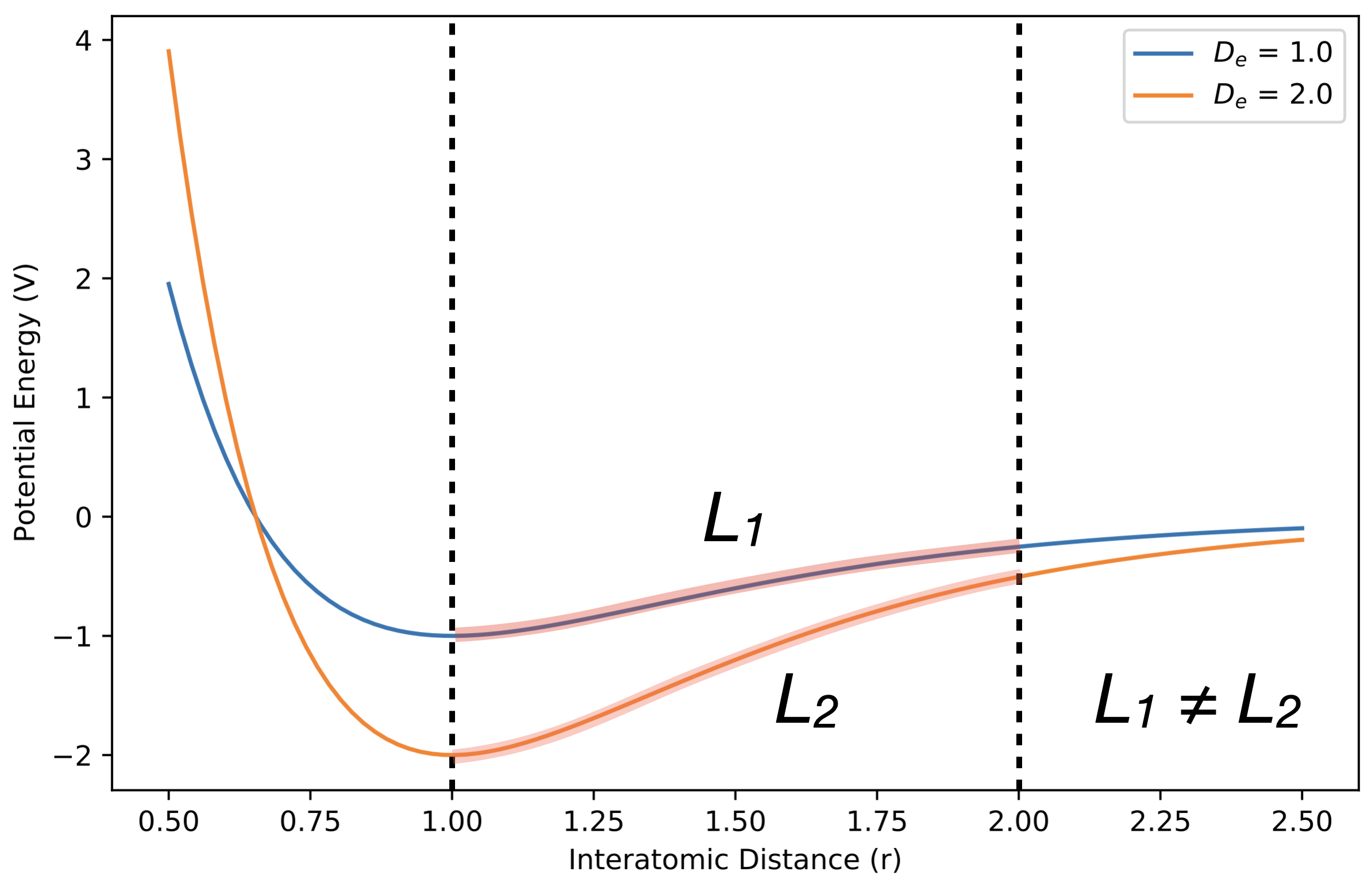}}
    \caption{The Morse potential, $V(r) = D_e\left(1 - e^{-a\left(r-r_e\right)}\right)^2-D_e$, for well-depth parameter $D_e=1,2$ for a simple two-atom model system. $r_e$ is the equilibrium distance between the two atoms. Despite the spatial similarity between structures at $r=1.0$ and $r=2.0$ being independent of the parameters of the potential, the geodesic distance between them changes as a consequence of changing the interaction, i.e. $L_1\rightarrow L_2$.}
    \label{fig:morse}
\end{figure}

Consider the heuristic example of the Morse potential that defines the interaction between two particles separated by a distance $r$ as
\begin{equation}
    V(r) = D_e\left( 1 - e^{-a(r-r_e)} \right)^2 - D_e,
    \label{eq:morse_potential}
\end{equation}
where the parameter $D_e$ governs the range of the interaction, $r_e$ defines the equilibrium length, and $a$ determines the decay rate of the interaction. Fig. \ref{fig:morse} shows the Morse interaction potential for two values of $D_e$.
As can be seen from the figure, two configurations, e.g., at $r=1.0$ and $r=2.0$, will be separated by the same Euclidean distance - or indeed any spatially-informed distance measure - regardless of the potential that operates on the structures. However, the \textit{geodesic distance along the potential energy manifold} between the two configurations is a unique consequence of the particular physical interaction (in this case, Morse potentials with $D_e = 1,2$) that operates on the structures. Referring to the ``Model = Representation + Architecture'' mnemonic in Section \ref{sec:introduction}, it can be seen that using a purely spatial, potential-invariant representation in a model that may be required to capture distinct types of physical interactions - e.g., attractive versus repulsive interactions - shifts much of the computational burden of adapting to the energy landscape on to the \textit{architecture}, since the representation is not adapted to the physics of the system. (Note that the short-/long-range interaction example is somewhat simplistic given the variety of solutions that have been proposed to address this problem. See, for instance, Ref. \citenum{long_range_interactions}.)

A principal argument in this work is that the geodesic distance measure has the capacity to efficiently encode information about the potential energy manifold that is absent in conventional spatial similarity measures, and that this information can be embedded into well-known descriptor schemes via appropriate parameterization. This argument is based on the following observations:
\begin{enumerate}
    \item \textbf{Geodesic distances are information-rich}: Given data about a discrete collection of structures and their energies (or forces), it is the task of interatomic potential models to make the best possible estimations of the energies of the continuum of structures that lie between - or beyond - those structures within a certain representation (descriptor) scheme. For this reason, one can also expect that any \textit{a priori} structure$\rightarrow$energy data about the continuum of states between these discrete structures is likely to benefit from a statistically optimized potential energy model. 
    While this information is not directly available in any limited, discrete set of training data, knowledge of geodesic distances (between a large number of pairs of structures) necessarily contains this information implicitly. Consider, for instance, two cases of the geodesic distance, $l_{AB}$, between structures $A$ and $B$, in relation to the direct, Euclidean distance, $l^0_{AB}$, between the structures, calculated as the length of the linear path in Cartesian coordinates between structures $A$ and $B$. \textbf{Case 1}: If $l_{AB} \approx l^0_{AB}$, this indicates the absence of a significant energy barrier or basin between the structures and, by extension, that any structures lying directly between them (in Cartesian coordinates) are unlikely to have very high potential energy compared to $A$ and $B$. \textbf{Case 2}: Conversely, if $l_{AB} \gg l^0_{AB}$, there are likely several high- or low-energy (compared to A and B) structures along the direct path between $A$ and $B$, due to which the minimum-energy path takes a more non-linear (thereby, longer) route between the structures. (Of course, $l_{AB}$ and $l^0_{AB}$ are intimately connected to the coordinate system that we utilize. We address this connection in more detail in Section \ref{sec:fast_approximation_geodesic_distances}.) If one could devise a way to capture this information directly within the descriptor scheme used to represent an atomistic system, this would - in principle - enable a fitting algorithm to access the rich geometric structure of the energy manifold, even if this is only approximately known.
    \item \textbf{Fast approximations exist for computing geodesic distances}: At first glance, it seems causally incorrect to expect that geodesic distances along the potential manifold will be available for parameterizing descriptors when the very purpose of the descriptors is to be able to approximate the potential that one does not know \textit{a priori} via some (learnable) function. 
    However, our goal is not to compute the exact geodesic distances between structures using the (theoretically) true potential energy function, but only to obtain reasonable approximations of geodesic paths that capture the general geometric features of the energy landscape. As we will discuss in Section \ref{sec:fast_approximation_geodesic_distances}, borrowing ideas from chemical reaction theory, it is possible to rapidly generate approximate geodesic paths and distances between a large number of pairs of structures at only a fraction of the cost of computing them with full \textit{ab initio} accuracy and using a pathway-search algorithm such as nudged elastic band.
    \item \textbf{Descriptor schemes contain several optimizable hyperparameters}: Popular schemes for generating atomic fingerprints often contain several hyperparameters that determine how an atomistic system is featurized. Beyond a fairly limited set of specifications regarding the range and scale of these values, they may be largely considered a set of optimizable hyperparameters that can be adjusted to best capture the relevant structural features of the system. Furthermore, descriptor schemes can utilize multiple copies of the same hyperparameter with different values to increase the structural ``completeness'' or uniqueness of the atomic fingerprint, introducing additional degrees of freedom that may be optimized or fine-tuned. Hyperparameter tuning generally follows a train/validation pipeline, wherein a subset of the overall dataset - termed the validation set - is used to evaluate, and possibly improve, a model's hyperparameters without being directly used to train the model. Instead, our thesis here is that the hyperparameters of a descriptor scheme can be optimized to maximize the descriptor's informativeness about the energy landscape \textit{without making reference to any particular model architecture}, purely through the use of information from pre-computable distance measures, such as geodesic distances. We discuss this in greater detail in the sections that follow, and demonstrate it in the context of ACSF descriptors.
\end{enumerate}
It is important to clarify here that geodesic distances are not necessarily always ``better'' or ``more informative'' than purely spatially-informed distance measures, such as Euclidean distance. Rather, our aim is to illustrate that, because geodesic distance incorporates a fundamentally different category of physical information - namely, energetic - than spatial distance, it is likely to contain at least some information that is beyond the capacity of spatial similarity measures to express. The goal then becomes twofold: First, to estimate the relative information content of two distance measures, e.g., Euclidean and geodesic. Second, to design a descriptor so as to maximize its information content given that different ground-truth distance measures may contain partially exclusive physical information.

We now discuss a potential strategy for the fast approximation of geodesic distances and its relevance for the design of more informative descriptors.

\subsection{Fast Approximation of Geodesic Distances and Application to Descriptor Design}
\label{sec:fast_approximation_geodesic_distances}
In the context of determining the minimum-energy (geodesic) pathways between equilibrium structures in reaction chemistry, it is of interest to generate initial guess pathways that are as accurate and close to the true minimum-energy pathways as possible. While early attempts at generating these guess pathways relied on naive approximations such as linear interpolation between the Cartesian coordinates of the initial and final structures,\cite{neb_original_paper} more advanced interpolation techniques have since been developed.

By assessing various interpolation schemes using a unifying differential-geometric approach, Zhu et al. (Ref. \citenum{geodesic_interpolation}) noted that the relative success of certain interpolation schemes - particularly those involving inverse bond length-based internal coordinate representations - over others, can be explained by the implicit choice of metric tensor (mapping length elements of the Riemannian manifold of the potential energy to the coordinate system) that accompanies these representations. More intuitively, they highlight that linear interpolation between two points (atomic configurations) in some system of coordinates is likely to resemble the geodesic path along the energy manifold if and only if the potential energy surface is ``almost flat'' when represented in those coordinates. Mathematically, given a potential energy function, $U$, and mass-weighted Cartesian coordinates, $\mathbf{x}$, the construction of an internal coordinate representation $\mathbf{q}$ such that
\begin{equation}
    \frac{\partial q^k}{\partial x_i}\frac{\partial q^k}{\partial x_j} \approx \frac{\partial U}{\partial x_i}\frac{\partial U}{\partial x_j}
    \label{eq:metric_equivalence}
\end{equation}
will necessarily result in straight-line paths in $\mathbf{q}$ approximating geodesic paths along the potential energy manifold. (Note that the Einstein summation convention is assumed in Eq. \ref{eq:metric_equivalence}.) 

An alternative way to state this would be to note that, the more closely an internal coordinate representation reflects the true geometry of the potential energy surface, the more accurately its Euclidean paths will correspond to geodesic paths along the manifold. They showed that it is possible to achieve this approximation using a simple system of internal coordinates given by
\begin{equation}
    q_i \equiv \exp \left[ -\alpha \frac{r_{kl} - r_e^{kl}}{r_e^{kl}}\right] + \beta\frac{r_e^{kl}}{r_{kl}},
    \label{eq:internal_coordinates}
\end{equation}
where $r_{kl}$ is the distance between atoms $k$ and $l$, and $r_e^{kl}$ is a parameter representing the equilibrium distance between $k$ and $l$. By adjusting the parameters $\alpha$ and $\beta$, it is possible to control the extent to which short- or long-range interactions operate between pairs of atoms given their separation; it is then straightforward to modify this scheme to include different parameters for chemically distinct pairs of atoms. Using this approach, they demonstrated that good approximate geodesic pathways between vastly different configurations may be constructed rapidly, even for biomolecular systems involving $\mathcal{O}(10^3)$ atoms. Furthermore, numerical approximations enable the geodesic distances between two configurations to be computed to arbitrary precision, which we do not discuss here for brevity.\\
\\
In the context of fitting physics-inspired potential energy forms to \textit{ab initio} data, descriptors are of particular significance due to the flexibility with which they can be designed to capture and emphasize various physical and chemical phenomena that one anticipates in a given class of atomistic systems. Descriptors are generally constructed in a way that captures the translation-, rotation-, and identical-atom permutation-invariance of the potential energy. As a consequence, many descriptors naturally fall under the definition of the internal coordinates referenced in the foregoing discussion.

This raises the question of why linear interpolation in the space of descriptors does not immediately result in geodesic interpolation along the manifold given that descriptors are designed to capture the features of potential energy surfaces. First, descriptors are not always constructed to satisfy the equivalence of their metric tensor with the Fukui metric for mass-weighted Cartesian coordinates, as depicted in Eq. \ref{eq:metric_equivalence}. If they were, this would largely obviate the need for highly non-linear fitting functions, such as neural networks, to model the energy landscape. Second, even if a potential manifold were overwhelmingly flat in the coordinate space of some descriptor, the repeated evaluation of analytical or numerical derivatives of these descriptors with respect to Cartesian coordinates - a requirement for the procedure of Ref. \citenum{geodesic_interpolation} - can be a highly non-trivial computational task that may be hard to generalize to new descriptor forms, except those explicitly designed to be used with gradient-based optimization algorithms.

Instead of attempting to construct descriptors that achieve the metric equivalence of Eq. \ref{eq:metric_equivalence}, we take an alternative approach. As described below, we use a data-driven procedure to encode the \textit{information content} of a large collection of easily computable pair-wise geodesic distances into the hyperparameters of descriptor forms without making any reference to the metric that governs their correspondence to the energy manifold. This allows us to retain the flexibility afforded by various descriptor schemes while utilizing additional information about the geometry of the potential energy landscape that can be determined prior to fitting the parameters of a statistical learning model to the training data, or indeed without even having defined such a model.

By using pre-computable data on the approximate geodesic distances between structures along the energy manifold we are able to endow the descriptor scheme with a more meaningful notion of how atomic configurations should be distributed in descriptor space. In making the underlying assumption that geodesic distances between structures - a quantity defined on the $(3N-6)$-dimensional energy manifold of an $N$-atom system - can by captured within the $\mathcal{O}$(1–10)-hyperparameter set of a descriptor, we are effectively invoking the manifold hypothesis\cite{gorban_manifold} before even passing the atomic fingerprints computed using the descriptor to a regression algorithm that attempts to fit its parameters to the energies. We note here that our descriptor optimization approach is not restricted to one particular method of computing geodesic distances, such as in Ref. \citenum{geodesic_interpolation}. It can be modified to accommodate other distance measures, enabling different types of information to be expressed. 

We demonstrate this procedure for atom-centered symmetry function (ACSF) descriptors (Section \ref{sec:acsf_descriptors}), which we have chosen due to their scalable local construction, physical interpretability in featurizing atomic environments, and widespread use.\cite{descriptors} We encode the information content within the descriptor hyperparameters using a Bayesian optimization algorithm (Section \ref{sec:theory_bayesian_optimization}) whose objective function is set to be the information imbalance between descriptor distances and a ground-truth distance measure, which we set to be the Euclidean and geodesic measures for comparison (Section \ref{sec:theory_information_imbalance}).


\subsection{Information Imbalance as an Objective Function}
\label{sec:theory_information_imbalance}
We now address the question of the choice of objective function to use for capturing the information content of a distance measure, e.g., the geodesic measure, in the descriptor's hyperparameters. First, we note that it is impractical to optimize the Euclidean distance between descriptor fingerprints to be identical to the geodesic distances between the structures themselves because the latter are unitless scalar quantities and there is no physical motivation for equalizing them. Instead, we propose using a statistical, information-theoretic quantity termed the symmetric information imbalance between two distance measures.\cite{distance_measures} This may be described briefly as follows. Consider two distance measures, $\mathcal A$ and $\mathcal B$. Given a collection of points distributed according to some unknown distribution, if we select any single point as a reference and rank each of the other points according to their distance from the reference point, it is possible that the rank ordering will not be identical between the two distance measures. This is equivalent to the observation that distances between points in two distinct distance measures need not vary monotonically with each other. The information imbalance in going from $\mathcal A$ to $\mathcal B$ is defined as
\begin{equation}
    \Delta (\mathcal A\rightarrow \mathcal B) = \frac{2\langle r_{ij}^\mathcal{B} | r_{ij}^\mathcal{A} = 1 \rangle_{ij}}{N},
    \label{eq:information_imbalance}
\end{equation}
where $r_{ij}^\mathcal{X}$ is the rank of point $j$ with respect to point $i$ in distance measure $\mathcal{X}$, and $N$ is the total number of points in the dataset. In essence, $\Delta (\mathcal A\rightarrow \mathcal B)$ answers the question, \textit{on average, for two points $i$ and $j$ that are nearest neighbors (rank $r_{ij}^A=1$) in distance measure $A$, what is their rank in distance measure B ($r_{ij}^B$)?} The \textit{symmetric} information imbalance is then defined as
\begin{equation}
    \bar{\Delta} (\mathcal A\leftrightarrow \mathcal B) = \frac{\Delta(\mathcal A\rightarrow \mathcal B) + \Delta(\mathcal B\rightarrow \mathcal A)}{\sqrt{2}},
    \label{eq:symmetric_information_imbalance}
\end{equation}
where $\bar{\Delta}$ simply symmetrizes the imbalance in going from one distance measure to the other.
In general, if $\Delta(\mathcal A\rightarrow \mathcal B) < \Delta (\mathcal B\rightarrow \mathcal A)$, distance measure $\mathcal A$ may be said to be more informative than $\mathcal B$, while a low value of $\bar{\Delta}(\mathcal A\leftrightarrow \mathcal B)$ indicates greater agreement between $\mathcal A$ and $\mathcal B$ than a higher (closer to 1) value.

Information imbalance has previously been applied to guide descriptor compression and has been used to assess the correspondence between distances in various descriptor representations and Euclidean distances in a dataset of Si atoms slightly perturbed from their bulk lattice positions.\cite{distance_measures} Interestingly, in Ref. \citenum{distance_measures}, the Euclidean distance is taken to be the ground-truth distance measure between structures by showing that increasingly expressive descriptors show lower information imbalance with respect to the Euclidean measure. As we will illustrate in Section \ref{sec:results_and_discussion_inequivalence}, however, the Euclidean and geodesic distance measures can fundamentally disagree. This, combined with the discussion in Section \ref{sec:theory_motivation}, suggests that there is no underlying reason for one of these contradictory distance measures to be chosen as the ground-truth over the other.

\subsection{Bayesian Optimization}
\label{sec:theory_bayesian_optimization}
Having defined an objective function, we formulate the optimization problem as follows. For a set of structures, we pre-compute all pair-wise distances, $\mathbf{r}^\mathcal{T}$, in the ground-truth (Euclidean or geodesic) distance measure, $\mathcal{T}$. Then, for a given set of descriptor hyperparameters $\mathbf p$, we evaluate the atomic fingerprints of each configuration in the structure set and compute the pair-wise distances, $\mathbf{r}^{\mathcal{D}(\mathbf p)}$ in the space of this descriptor, $\mathcal{D}(\mathbf p)$. We then evaluate the objective function, i.e., the symmetric information imbalance between the descriptor and the ground-truth measure given by
\begin{equation}
    \bar{\Delta}(\mathcal{D}(\mathbf{p})\leftrightarrow \mathcal{T}) = \frac{\sqrt{2}}{N} \left(\langle r^{\mathcal{D}(\mathbf{p})}_{ij} | r^\mathcal{T}_{ij}=1\rangle_{ij} + \langle r^\mathcal{T}_{ij} | r^{\mathcal{D}(\mathbf{p})}_{ij}=1\rangle_{ij} \right).
    \label{eq:objective_function}
\end{equation}
The task of the optimization algorithm then becomes to iteratively update the set of hyperparameters $\mathbf p$ in order to identify the set that minimizes the information imbalance of the descriptor distance measure with respect to the ground-truth measure.

The question now arises of which optimization algorithm to use. Since the objective function is a statistical quantity, it is difficult to rely on gradient-based algorithms, such as gradient descent. Moreover, since new hyperparameter sets must be generated and passed to Eq. \ref{eq:objective_function} on-the-fly, it is desirable to use an algorithm that operates in data-sparse regimes and intelligently queries new data points for inexpensive, data-efficient optimization. Computing the information imbalance from the pair-wise distances of a large set of structures is a moderately expensive computational operation, which further encourages the use of such an efficient algorithm. As we discuss below, Bayesian optimization presents an elegant solution to these problems.

Bayesian optimization has emerged as a powerful strategy for the gradient-free, stochastic optimization of black-box function parameters. Its data efficiency, coupled with the ability to systematically explore the landscape of the objective function through a probabilistic framework, makes it particularly appealing for tasks in which function evaluations are computationally expensive. This optimization technique leverages a probabilistic model - typically a Gaussian process - to model the objective function and utilizes an acquisition function to guide the search for the optimum by balancing exploration and exploitation.\cite{bayes_opt_review} This enables the algorithm to incorporate prior knowledge and iteratively update it with new observations, thereby refining its model of the objective function with each iteration. Such an approach is in stark contrast to grid or random search methods, which lack the ability to learn from prior evaluations and can be prohibitively expensive in high-dimensional spaces or when evaluations are costly.\cite{practical_bayes_opt}

In mainstream machine learning, Bayesian optimization has been successfully applied to the task of hyperparameter tuning. Traditional methods for hyperparameter optimization, such as exhaustive search, are often impractical due to the exponential growth of the search space with the number of hyperparameters. Bayesian optimization addresses this challenge by constructing a probabilistic model of the performance function and iteratively selecting the most promising hyperparameters to evaluate based on the current model. This approach has been shown to significantly reduce the number of evaluations needed to identify optimal or near-optimal hyperparameters.\cite{bayes_hyperparameter_opt}

In the present context, we use Bayesian optimization to intelligently query new descriptor hyperparameter sets and compute the information imbalance between that descriptor scheme and the ground-truth distance measure, e.g., the Euclidean or geodesic measure.

\section{Computational Details}
\label{sec:computational_details}
\subsection{ACSF Descriptors}
\label{sec:acsf_descriptors}
Atom-centered symmetry function (ACSF) descriptors rely on the construction of functions centered around each atom ($i$) of a system by featurizing only the local neighborhood of that atom.\cite{behler, behler_parrinello} Here, we begin by focusing on $G^\textrm{II}$- and $G^\textrm{IV}$-type ACSF descriptors, which take the form
\begin{equation}
    G_i^\textrm{II} = \sum_{j\neq i}^{R_{ij}<R_c} e^{-\eta\left(R_{ij}-R_s\right)^2/R_c^2}f_c\left(R_{ij}\right),
    \label{eq:acsf_g2}
\end{equation}
and
\begin{equation}
    G_i^\textrm{IV} = 2^{1-\zeta}\sum_{j,k\neq i}^{R_{ij},R_{ik}<R_c} \left( 1 + \lambda\cos\theta_{ijk} \right)^\zeta e^{-\eta\left( R_{ij}^2+R_{jk}^2+R_{ki}^2 \right)} f_c\left(R_{ij}\right) f_c\left(R_{jk}\right) f_c\left(R_{ki}\right).
    \label{eq:acsf_g4}
\end{equation}
In Eqs. \ref{eq:acsf_g2} and \ref{eq:acsf_g4}, $R_{ij}$ is the separation between atoms $i$ and $j$, $R_c$ is a cutoff radius parameter, $R_s$ is a small displacement to the center of the spherical shell around which the descriptor is constructed, $\eta$, $\zeta$, and $\lambda$ are tunable parameters, and $f_c(\cdot)$ is a smooth cosine-type tapering function that goes to 0 at $R_c$.

We now describe the three schemes of ACSF descriptors that we optimize using our approach.
\begin{enumerate}
    \item \textbf{Scheme I}: $G^\textrm{II}$-type descriptors with 2 $\eta$ parameters,
    \item \textbf{Scheme II}: $G^\textrm{II}$-type descriptors with 4 $\eta$ parameters, and 
    \item \textbf{Scheme III}: $G^\textrm{II}$- and $G^\textrm{IV}$-type descriptors with 4 $\eta$ parameters in the $G^\textrm{II}$ descriptor, 1 (different)  $\eta$ parameter in the $G^\textrm{IV}$ descriptor, 1 $\zeta$ parameter in the $G^\textrm{IV}$ desciptor, and 2 $\lambda$ parameters in the $G^\textrm{IV}$ descriptor.
\end{enumerate}

\subsection{Optimization Settings}
\label{sec:optimization_parameters}
The hyperparameters of each descriptor defined in Section \ref{sec:acsf_descriptors} are then treated as the variables that we attempt to optimize with the Bayesian optimization algorithm. We perform all Bayesian optimization calculations using the \texttt{scikit-optimize} package.\cite{scikit_optimize} For each loop of the algorithm, we use 100 function calls and 50 random initializations, which we found to be sufficient for minimizing the information imbalance given the dimensionality of the hyperparameter space. The information imbalance with respect to a given ground-truth distance measure - whether Euclidean or geodesic - was computed in each case using in-house Python implementations of Eqs. \ref{eq:information_imbalance}–\ref{eq:objective_function}.\cite{distance_measures} At the end of an optimization cycle, the hyperparameter set that achieves the lowest value of the objective function is chosen as the optimized set. If more than one hyperparameter set achieves an optimal value of the objective function, then one of these sets is chosen at random by the algorithm, which we report directly.

Further details on the constrained optimization of the hyperparameters are provided in Section \ref{sec:results_bayesian_optimization}.

\subsection{Dataset and Ground-Truth Distance Computation}
We perform our analysis using molecular configurations of the ethanol, malonaldehyde, and aspirin datasets of the MD22 database.\cite{md22} These datasets include forces and energies computed using coupled-cluster theory with single, double, and perturbative triple excitations (CCSD(T)) on atomic configurations taken from \textit{ab initio} molecular dynamics (AIMD) trajectories. These AIMD simulations themselves utilize DFT-PBE calculations with van der Waals corrections and are performed at 400–500 K.\cite{md22} The use of high-temperature MD simulations is significant because it indicates a higher likelihood of sufficiently broad sampling of the energy landscape including the traversal and crossing of potential energy barriers. We choose these molecules as test cases due to the conformational flexibility observed in the molecular dynamics trajectories of these molecules in the MD22 database. Furthermore, the correct measurement of Euclidean distances and geodesic distances between different configurations first requires them to be maximally aligned, both translationally and rotationally, prior to evaluating the displacement of each of the individual atoms and the geodesic pathways between configurations. The relatively small size of these molecules makes such alignment fairly unambiguous - in this case, by first aligning their center-of-mass coordinates and then using the Kabsch algorithm\cite{kabsch}.

Euclidean distances are computed in the usual way, i.e., as the $L^2$ norm of the vector difference between the Cartesian representation of two structures. Geodesic distances are computed as in Ref. \citenum{geodesic_interpolation} with the values of the parameters $\alpha$ and $\beta$ in Eq. \ref{eq:internal_coordinates} being set to 1.7 and 0.01, respectively, in all cases. We note, however, that it might even be useful to calibrate these parameters using energies from the training dataset prior to computing geodesic distances. We compute Euclidean and geodesic distances between all pairs of structures from a subset of 100 structures from the dataset of each molecule, chosen using a farthest-point sampling algorithm.\cite{imbalzano_ceriotti_sampling_2018}

\section{Results and Discussion}
\label{sec:results_and_discussion}
\subsection{Inequivalence of Euclidean and Geodesic Distance Measures}
\label{sec:results_and_discussion_inequivalence}
\begin{figure}[]
    \centering
    \begin{subfigure}[b]{0.49\textwidth}{\includegraphics[width=\textwidth]{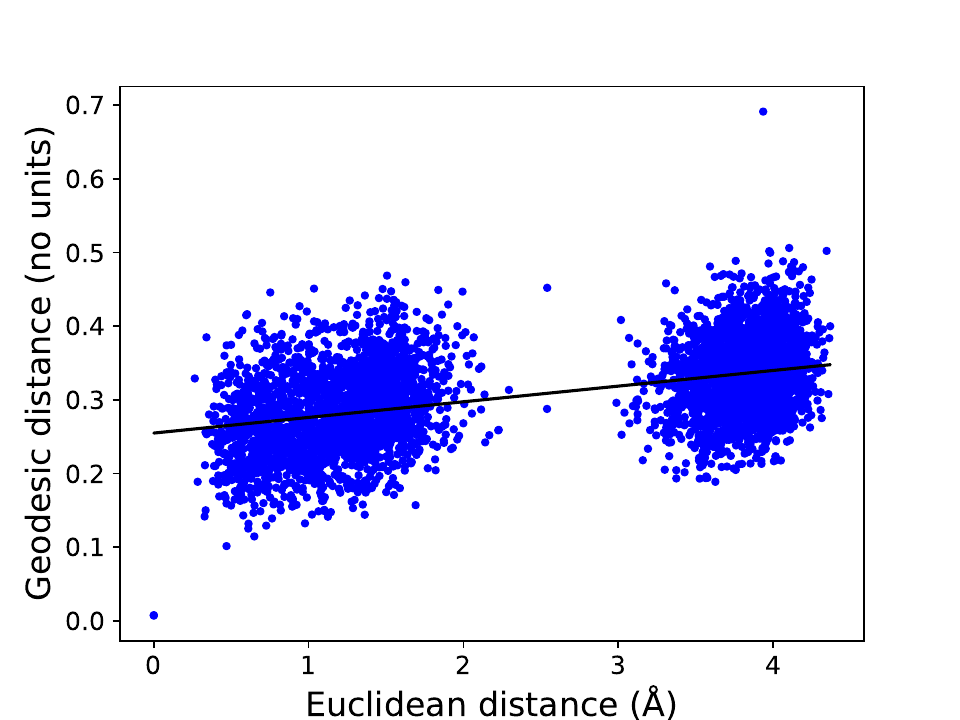}}
    \caption{}
    \label{fig:non_monotonicity}
    \end{subfigure}
    \hfill
    \begin{subfigure}[b]{0.49\textwidth}{\includegraphics[width=\textwidth]{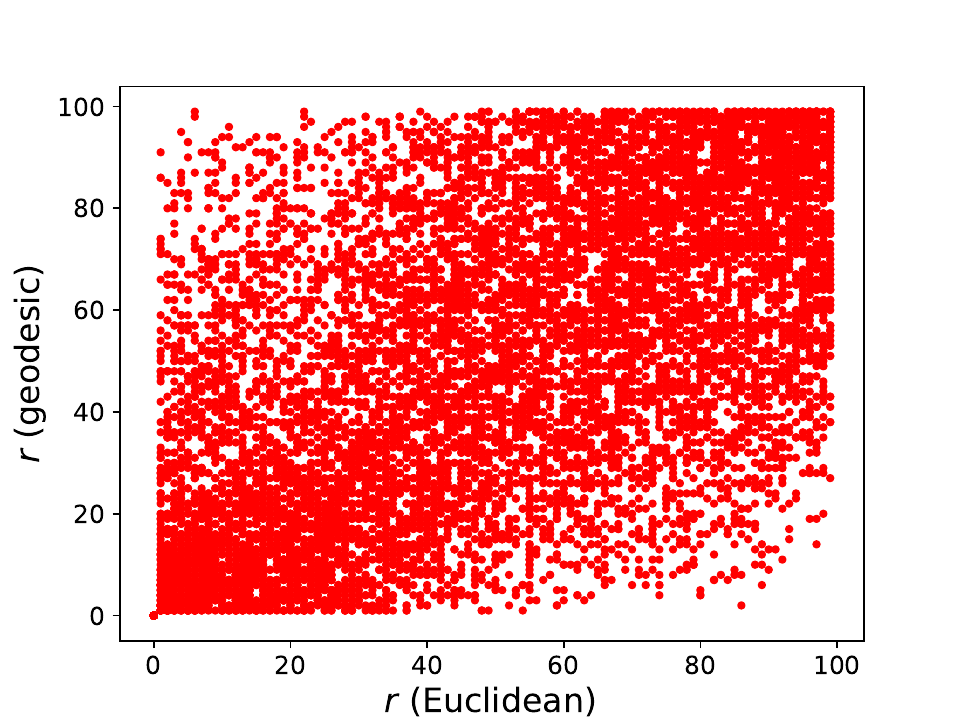}}
    \caption{}
    \label{fig:rank_imbalance}
    \end{subfigure}
    \caption{Non-equivalence of Euclidean and geodesic distance measures demonstrated on all pairwise distances in a set of 100 structures from the MD22 ethanol training dataset. (a) Non-monotonicity between Euclidean and geodesic distances. Despite the general collective increase of geodesic distance with increasing Euclidean distance as seen by the positive slope of the black regression line, the relationship is much more ambiguous at the level of individual configurations. (b) Scatter plot of ranks between pairs of points in both distance measures.}
    \label{fig:ethanol_demonstration}
\end{figure}

First, we provide a more concrete demonstration of the inequivalence of the Euclidean and geodesic distances via the example of pair-wise distances in a 100-structure subset of the ethanol training dataset of the MD22 database. In Fig. \ref{fig:non_monotonicity}, we clearly observe that the geodesic distances vary highly non-monotonically with the Euclidean distance, with only a slight positive correlation emerging when many - i.e., ${100\choose2}$ in this case - pairs are considered. It can also be observed that there is a greater spread of the Euclidean distances than the geodesic distances, suggesting that pairs of configurations may seem more different from each other based on the Euclidean distance despite being relatively similar based on the geodesic distance. This shows that the Euclidean distance alone can be somewhat misleading as a metric for the separation between configurations if one wants to optimally sample configurations along a potential energy landscape. Now, plotting the ranks (indexed closeness) of every point with respect to every other point in the data subset using the geodesic distance measure versus the Euclidean distance measure yields Fig. \ref{fig:rank_imbalance}. Given that distance measures in good agreement should display an almost linear relationship in such a parity plot, the highly noisy nature of Fig. \ref{fig:rank_imbalance} indicates generally poor agreement between the Euclidean and geodesic distance measures. Equivalent plots for malonaldehyde and aspirin are shown in Fig. \ref{fig:malonaldehyde_and_aspirin_demonstration} of the Supporting Information.

\begin{figure}[]
    \centering
    {\includegraphics[width=0.72\textwidth]{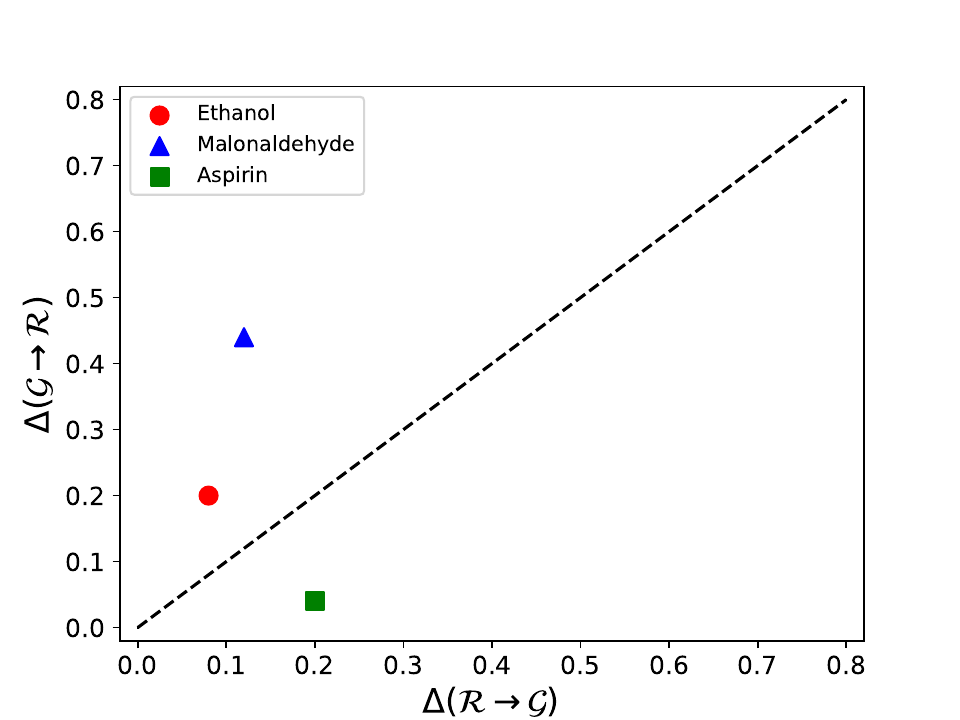}}
    \caption{Information imbalance between Euclidean and geodesic distances measured on all pairwise distances in a set of 100 structures from the MD22 training datasets of ethanol, malonaldehyde, and aspirin. $\Delta (\mathcal{R}\rightarrow\mathcal{G})$ refers to the information imbalance in going from the Euclidean measure ($\mathcal{R}$) to the geodesic measure ($\mathcal{G}$), and vice versa. The dashed line ($y=x$) broadly divides the plot into regions of relative informativeness. For example, regions above the dashed line indicate that $\mathcal R$ contains the information in $\mathcal G$ and vice versa.}
    \label{fig:imbalance_demonstration}
\end{figure}

Next, we demonstrate this more quantitatively by plotting the information imbalance in going from the geodesic distance measure ($\mathcal G$) to the Euclidean distance measure ($\mathcal R$), i.e., $\Delta(\mathcal G \rightarrow \mathcal R)$, versus the reverse, $\Delta (\mathcal R \rightarrow \mathcal G)$. This is depicted in Fig. \ref{fig:imbalance_demonstration} for a 100-structure subset of the ethanol, malonaldehyde, and aspirin training datasets of the MD22 database. Recall that, in this representation, points above the $y=x$ line indicate that the Euclidean distance measure is ``more informative'' than the geodesic distance measure and vice versa, while a value that does not lie close to the origin indicates a significant, finite disagreement between the two distance measures.

It can be seen from Fig. \ref{fig:imbalance_demonstration} that the Euclidean and geodesic distance measures are not necessarily in good agreement. Conceptually, Fig. \ref{fig:imbalance_demonstration} may be interpreted as follows. If a point lies above (below) the dashed line $y=x$, then any two atomic configurations in the dataset that are separated by a small Euclidean (geodesic) distance are also statistically likely to also be separated by a small geodesic (Euclidean) distance. However, the converse is less true, i.e., two configurations separated by a small geodesic (Euclidean) distance are statistically \textit{less likely} to also be separated by a small Euclidean (geodesic) distance. An alternative way to state this would be to say that low Euclidean (geodesic) distance is a better indicator of low geodesic (Euclidean) distance than the reverse. In more concrete, physical terms, this interpretation may be stated in the context of potential energy manifolds following the discussion in Section \ref{sec:theory_motivation}. Consider two atomic configurations $A$ and $B$ separated by a moderate geodesic distance. Of several possibilities for the geodesic path connecting $A$ and $B$, consider the following two cases: (1) $A$ and $B$ are spatially distant but are separated by a smooth, flat region on the potential energy surface due to which the geodesic distance is directly correlated with the linear (Euclidean) path between the structures; and (2) $A$ and $B$ are spatially close but are separated by significant energy barriers or basins due to which the direct Euclidean path does not accurately reflect the geodesic path. This entails a possible non-uniqueness in the mapping from geodesic distance to Euclidean distance and vice-versa. Consider the case where the mapping from geodesic distance to Euclidean distance is less unique (more degenerate) than the inverse map, i.e., $\Delta(\mathcal R\rightarrow \mathcal G) < \Delta(\mathcal G\rightarrow \mathcal R)$. This suggests that a small spatial (Euclidean) separation between configurations $A$ and $B$ is a better indicator of geodesic closeness than a small geodesic distance between $A$ and $B$ is an indicator of Euclidean closeness.

Of course, as seen in Fig. \ref{fig:imbalance_demonstration}, the extent to which this (or its opposite) is true depends on the chemistry of the specific system being studied. For example, geodesic distances appear more informative for the aspirin sub-dataset, while Euclidean distances appear more informative (to varying degrees) for the ethanol and malonaldehyde sub-datasets. We note that providing mechanistic explanations for the observed values of information imbalance is a non-trivial task. This is because information imbalance is a statistical quantity computed over many pairs of structures that may display a wide range of behaviors in the relationship between Euclidean and geodesic distances. Instead, our goal with these examples is primarily to show that Euclidean and geodesic distances between atomic configurations tend to be inequivalent, as evidenced by the non-uniqueness in the mapping from one distance measure to the other. 

We now discuss the optimization of the ACSF descriptors with the objective of minimizing the symmetric information imbalance between distances in descriptor space and one or both of the Euclidean and geodesic distance measures.

\subsection{Combined-Objective Descriptor Optimization}
\label{sec:results_bayesian_optimization}
The descriptor schemes defined in Section \ref{sec:acsf_descriptors} contain few to several hyperparameters which can be treated as the optimizable parameters for the Bayesian optimization framework. However, indiscriminately optimizing them not only increases the burden on the optimizer, it may also wastefully iterate through many meaningless or unphysical functions. To avoid this, we optimize the (hyper)parameters of the descriptors with the following constraints:
\begin{enumerate}
    \item \textbf{Scheme I}: First, we optimize $R_c$ and $R_s$ restricting them to the continuous range of $(1,7)$ Å. The two $\eta$ parameters, $\eta_1$ and $\eta_2$, are optimized separately. We first optimize two auxiliary parameters, $x_1\in[\epsilon, 1-\epsilon]$ and $x_2\in[1+\epsilon,20]$, where $\epsilon$ is some small value $\sim10^{-6}$, and subsequently set $\eta_1 = \log_{10}(x_1)$ and $\eta_2 = \log_{10}(x_2)$.
    \item \textbf{Scheme II}: Similar to Scheme I, in addition to $R_c$ and $R_s$, two parameters $x_1$ and $x_2$ are first optimized, following which four $\eta$ parameters are defined to lie evenly spaced on the $\log_{10}$ scale, with the endpoints being set to $\eta_1 = \log_{10}(x_1)$ and $\eta_4 = \log_{10}(x_4)$.
    \item \textbf{Scheme III}: In addition to $R_c$, $R_s$, and the four $\eta$ parameters optimized according to Scheme II, we optimize a single $\eta$ parameter for the $G^\textrm{IV}$ descriptors. Further, we optimize $\zeta\in[2,64]$, requiring that it is an integer and explored on the log scale. $\lambda$ is also set to $-1$ and $1$ alternately.
\end{enumerate}

Using each of these schemes, we now utilize the Bayesian optimization algorithm described in Section \ref{sec:theory_bayesian_optimization} to optimize the descriptor hyperparameters so as to minimize the symmetric information imbalance between the \textit{Euclidean distance measure in descriptor space}, $\mathcal D$, and either the Euclidean distance measure (in the original Cartesian coordinates), $\mathcal R$, or the geodesic distance measure, $\mathcal G$. This results in two different objective functions, $\bar{\Delta}(\mathcal D \leftrightarrow \mathcal R)$ and $\bar{\Delta}(\mathcal D \leftrightarrow \mathcal G)$, with which the descriptor can be designed. However, to demonstrate that it is possible for a single descriptor to reconcile the unshared information contained in both the Euclidean and geodesic distance measures, we also optimize the descriptors to minimize a third objective function, which we define to be the sum of the two previous information imbalances, i.e., $\bar{\Delta}(\mathcal D \leftrightarrow \mathcal R) + \bar{\Delta}(\mathcal D \leftrightarrow \mathcal G)$.

\begin{figure}[]
    \centering
    \begin{subfigure}[b]{0.49\textwidth}{\includegraphics[width=\textwidth]{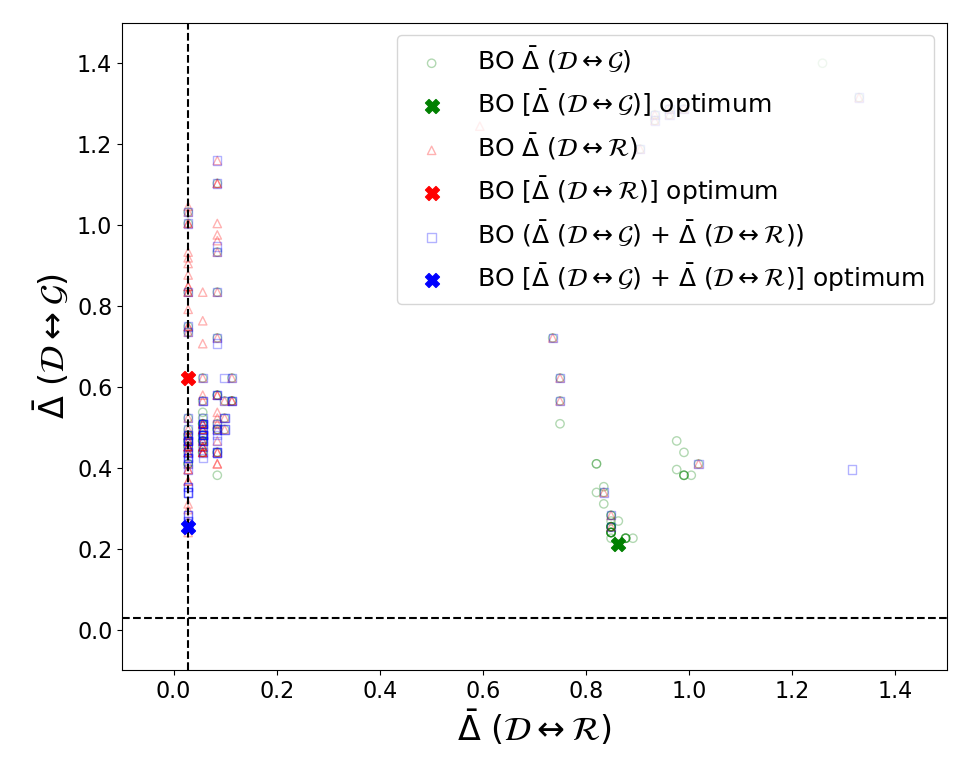}}
    \caption{}
    \label{fig:BO_ethanol}
    \end{subfigure}
    \hfill
    \begin{subfigure}[b]{0.49\textwidth}{\includegraphics[width=\textwidth]{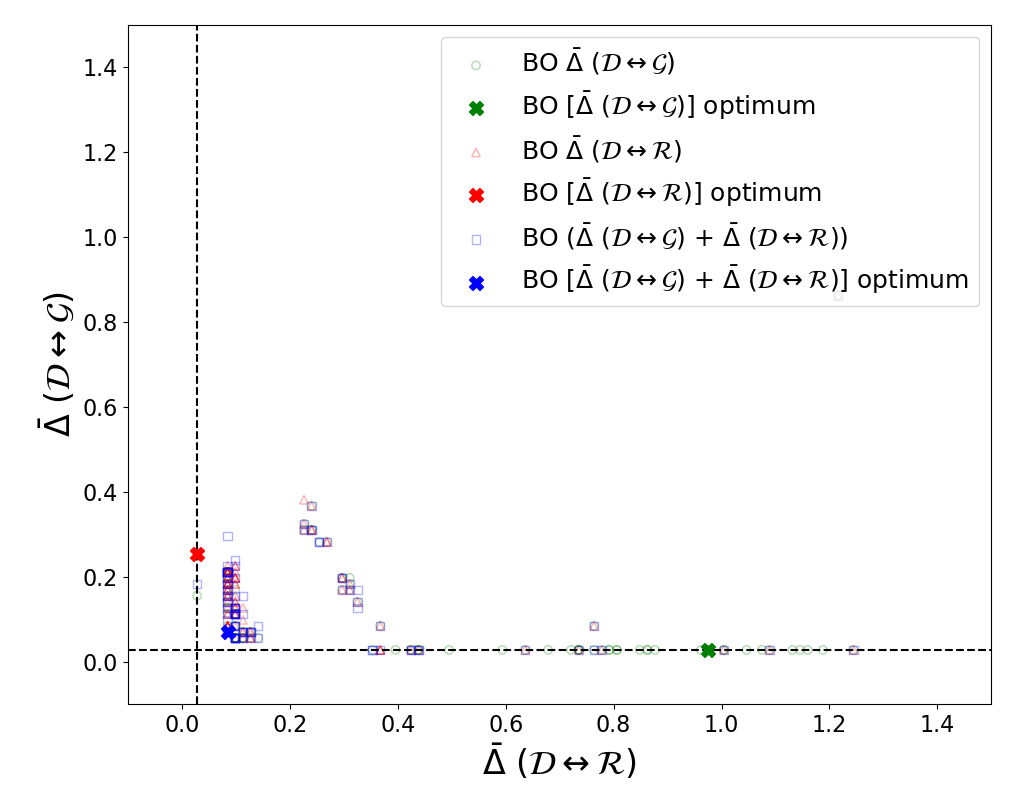}}
    \caption{}
    \label{fig:BO_malonaldehyde}
    \end{subfigure}
    \hfill
    \begin{subfigure}[b]{0.49\textwidth}{\includegraphics[width=\textwidth]{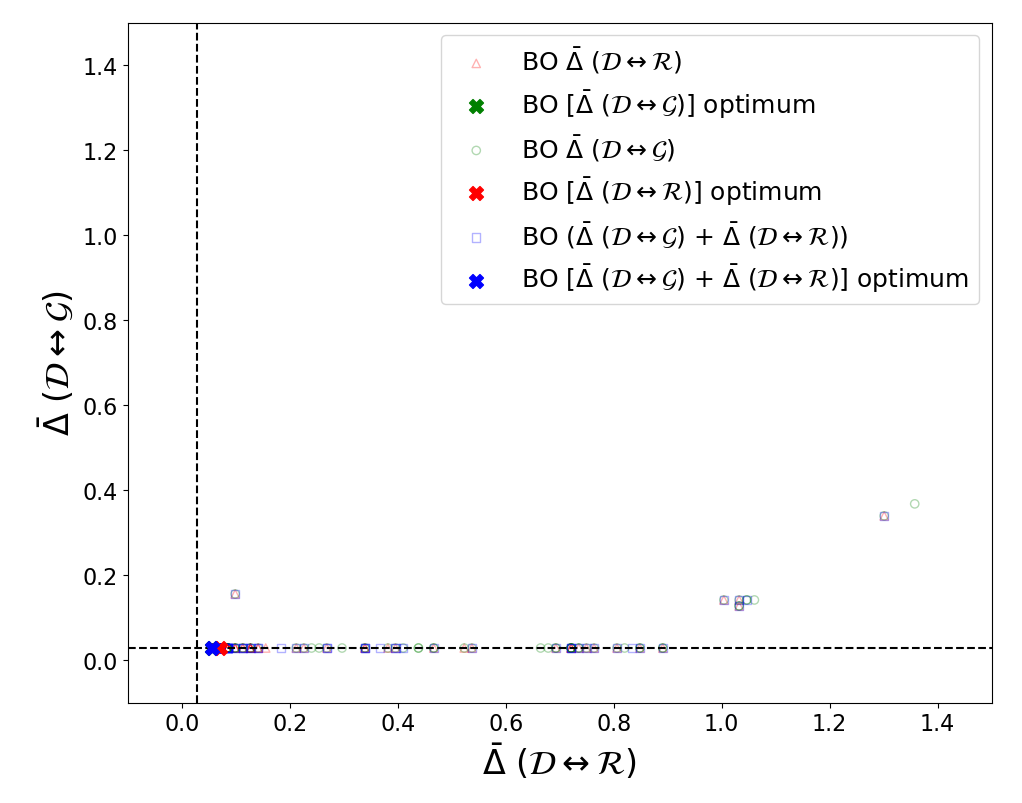}}
    \caption{}
    \label{fig:BO_aspirin}
    \end{subfigure}
    \caption{Bayesian optimization samples of descriptor parameters (Scheme 3) for (a) ethanol, (b) malonaldehyde, and (c) aspirin. The objective function is chosen to minimize the symmetric information imbalance between the descriptor distance measure ($\mathcal{D}$) and the geodesic distance ($\mathcal{G}$) [i.e. $\bar{\Delta}(\mathcal{D}\leftrightarrow\mathcal{G})$, green circles], the Euclidean distance ($\mathcal{R}$) [i.e. $\bar{\Delta}(\mathcal{D}\leftrightarrow\mathcal{R})$, red triangles], and the sum of both imbalances [i.e. $\bar{\Delta}(\mathcal{D}\leftrightarrow\mathcal{G}) + \bar{\Delta}(\mathcal{D}\leftrightarrow\mathcal{R})$, blue squares]. Crosses represent the optimal descriptor parameters suggested at the end of each Bayesian optimization loop. Combined optimization leads to descriptors closest to the origin (blue crosses), i.e., encoding information about both ground-truth distance measures. Black dashed lines mark the theoretical limit for 100 samples.}
    \label{fig:bayesian_optimization}
\end{figure}

In Fig. \ref{fig:bayesian_optimization}, we show the results of Bayesian optimization of each of these objectives as plots of $\bar{\Delta}(\mathcal D \leftrightarrow \mathcal R)$ versus $\bar{\Delta}(\mathcal D \leftrightarrow \mathcal G)$ for each of three systems - ethanol, malonaldehyde, and aspirin - for descriptor Scheme 3. Similar plots for Schemes 1 and 2 are shown in Figs. \ref{fig:supp_bayesian_optimization_1} and \ref{fig:supp_bayesian_optimization_2} of the Supporting Information. We note that, in each of these plots, the closer a point (i.e., an optimized descriptor) lies to the origin the better, with the best descriptors lining the Pareto frontier of the plot. In Bayesian optimization, if multiple explored points optimize the objective equally well, they are considered equivalent and any single one of them may be chosen and reported by the algorithm as the optimal choice. (We do, in fact, observe this phenomenon frequently when the optimizer is tasked with minimizing either $\bar{\Delta}(\mathcal D\rightarrow \mathcal G)$ or $\bar{\Delta}(\mathcal D\rightarrow \mathcal R)$.) Fig. \ref{fig:bayesian_optimization} shows that - notably in the case of ethanol and malonaldehyde - given the objective of minimizing only $\bar{\Delta}(\mathcal D \leftrightarrow \mathcal R)$, Bayesian optimization identifies descriptors that capture the information in the Euclidean distance measure reasonably well, albeit without much shared information with the geodesic measure. The reverse is also found to be true when optimizing only $\bar{\Delta}(\mathcal D \leftrightarrow \mathcal G)$. However, the use of the combined objective function, $\bar{\Delta}(\mathcal D \leftrightarrow \mathcal R) + \bar{\Delta}(\mathcal D \leftrightarrow \mathcal G)$, generates descriptors that optimally balance the information content contained in both the Euclidean and geodesic distance measures, as seen by the bold blue cross points, which are consistently the closest to the origin. Comparing Figs. \ref{fig:bayesian_optimization}, \ref{fig:supp_bayesian_optimization_1}, and \ref{fig:supp_bayesian_optimization_2}, it is also interesting to note that this theme is observed regardless of the number of hyperparameters in the descriptor scheme.

\begin{figure}[]
    \centering
    {\includegraphics[width=1.0\textwidth]{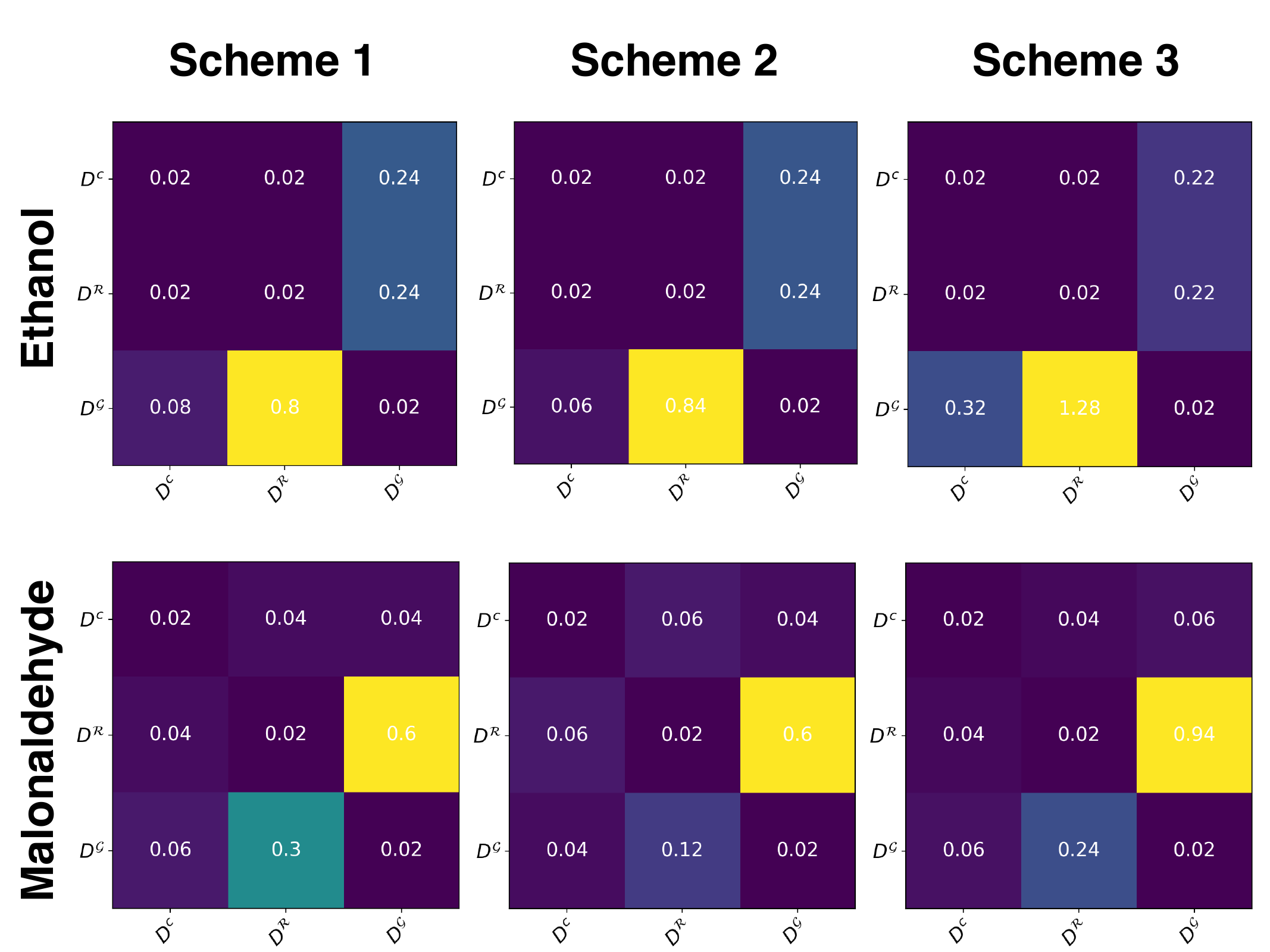}}
    \caption{Information imbalance between three types of descriptors. $D^\mathcal{R}$ is optimized to minimize the symmetric information imbalance ($\bar{\Delta}$) between descriptor distances and Euclidean distances. $D^\mathcal{G}$ is optimized to minimize $\bar{\Delta}$ between descriptor distances and geodesic distances. $D^c$ is optimized to minimize the sum of symmetric information imbalances between descriptor distances and both Euclidean and geodesic distances. In each matrix, entry $(i,j) = \Delta(D^i\rightarrow D^j)$. The confinement of large imbalance values ($>$ 0.5) to the $D^\mathcal{R}\times D^\mathcal{G}$ sub-matrices indicates that the joint Euclidean–geodesic optimization of descriptors leads them to contain more comprehensive distance information.}
    \label{fig:heatmaps}
\end{figure}

Given these newly constructed descriptors, it is now desirable to check whether descriptors constructed by minimizing one objective, say $\bar{\Delta}(\mathcal D \leftrightarrow \mathcal R)$, are more, less, or equally informative compared to descriptors constructed by minimizing the other objectives. We present this in Fig. \ref{fig:heatmaps}. In these heatmaps, $D^c$, $D^\mathcal{R}$, and $D^\mathcal{G}$ refer to descriptors that are optimized by minimizing (a) the combination of Euclidean and geodesic information imbalances, $\bar{\Delta}(\mathcal D \leftrightarrow \mathcal R) + \bar{\Delta}(\mathcal D \leftrightarrow \mathcal G)$; (b) only the Euclidean imbalance, $\bar{\Delta}(\mathcal D \leftrightarrow \mathcal R)$; and (c) only the geodesic imbalance, $\bar{\Delta}(\mathcal D \leftrightarrow \mathcal G)$. The entries of each matrix in Fig. \ref{fig:heatmaps} are arranged as follows: $(i, j) = \Delta(D^i\rightarrow D^j)$, where we now note the \textit{asymmetric} information imbalance.

In the case of ethanol and malonaldehyde, we note that high ($> 0.5$) values of information imbalance - in some cases, significantly high, $\sim1$ - are restricted to the $D^\mathcal{R}\times D^\mathcal{G}$ sub-block of each of the $3\times3$ matrices. We do not include the results for aspirin since all descriptors are found to be in complete agreement with each other in this case. In the case of ethanol, the highest of these correspond to the imbalance term $\Delta(D^\mathcal{G}\rightarrow D^\mathcal{R})$, whereas for malonaldehyde, these correspond to $\Delta(D^\mathcal{R}\rightarrow D^\mathcal{G})$. This highlights the unintuitive fact that \textit{substantial disagreements in distance information between two similarly constructed descriptors can arise from the seemingly inconsequential fact that their hyperparameters better encode one type of ground-truth distance information over the other}. Conversely, optimizing the descriptors to minimize the combined objective leads to better agreement with descriptors constructed using only one objective, indicating that combined-objective optimization is capable of resolving the disagreement between contradictory structural representations.

\section{Conclusions and Outlook}
\label{sec:conclusions}
In this work, we have proposed a novel paradigm for optimizing and evaluating atomistic descriptors by taking into account the partially exclusive information content of Euclidean and geodesic distance measures in molecular configuration space. We have presented arguments in favor of utilizing updated ground-truth distance measures that account for physical information beyond spatial similarity, demonstrating this via the fundamental disagreements between Euclidean and geodesic distances. We have then introduced the relevance of this approach in atomistic descriptor design through a Bayesian optimization algorithm that utilizes information imbalance as an objective function, thereby showing that it is possible to construct descriptors that maximize their shared information with a given distance measure, either Euclidean or geodesic. We have also shown that disagreements between descriptors that encode distances in only one distance measure can be resolved by a combined-objective Bayesian optimization strategy. This is noteworthy because, by minimizing the information imbalance between a descriptor scheme and conflicting distance measures, we maximize the uniqueness of atomic fingerprints generated using that descriptor. This, in turn, makes them more amenable to unambiguous sampling - structural or energetic - and statistical modeling.\cite{incompleteness_1, incompleteness_2}

Atomistic representations in which direct (linear) distances correspond to geodesic distances along potential energy manifolds may have the capacity to disentangle the often highly complex, nonlinear structure of these manifolds, rendering them tractable to easier interpretation, partitioning (into similar and dissimilar regions), and fitting by statistical learning models. Given that geodesic distances represent a fundamentally different category of physical information in addition to the structural information that is available from an atomic configuration, this approach enables us to encode some approximate knowledge of the potential energy surface \textit{a priori} into the hyperparameters of the descriptor, thereby maximizing the physical informativeness of an otherwise limited, closed-form structural representation of the atomistic system.

Given the utility of machine learning-based interatomic potentials, the application of our approach to an atomistic machine learning task would constitute a compelling demonstration of its utility. However, given the extensive scope for design, testing, and analysis this entails, we leave this avenue of exploration for subsequent work. Our goal with this work has been to introduce a unique way of thinking about atomistic representations and the modeling of potential energy landscapes. One may extend this concept to the inclusion of yet other types of information, besides structural and energetic, for instance, by treating state likelihood or prediction uncertainty as a physical quantity that can be encoded into descriptor representations. This may be considered similar in spirit to metadynamics\cite{metadynamics} or uncertainty-driven methods that have been previously suggested in the context of active learning for atomistic machine learning potentials.\cite{udd_active_learning}

\section{Data Availability Statement}

The code necessary to reproduce the results of this study is openly available online at \texttt{https://github.com/gopal-iyer/geodesic\_descriptor\_optimization/tree/main}.

\section{Supporting Information}
The Supporting Information document contains plots demonstrating (1) the non-equivalence of Euclidean and geodesic distance measures in the examples of malonaldehyde and aspirin; and (2) plots depicting Bayesian optimization samples for optimizing the hyperparameters of descriptor Schemes 2 and 3.

\section{Disclosure Statement}

The authors have no competing interests to declare. 

\begin{acknowledgement}
G.R.I. (concept, derivation, implementation, drafting of the manuscript) was funded by AFOSR Award Number FA9550-19-1-9999 and the Brown University Chemistry Department Vince Wernig Fellowship. B.R. (concept, drafting of the manuscript) was funded by NSF CTMC CAREER Award 2046744. This research was conducted using computational resources and services at the Center for Computation and Visualization, Brown University.
\end{acknowledgement}

\bibliography{main}


\end{document}